\documentclass[prl,twocolumn,showpacs,preprintnumbers,amsmath,amssymb]{revtex4}
\usepackage{graphicx}% Include figure files
\usepackage{dcolumn}% Align table columns on decimal point
\usepackage{bm}% bold math
\usepackage{psfig}
\usepackage{epsfig}

\begin{document}

\newcommand{\vAi}{{\cal A}_{i_1\cdots i_n}}
\newcommand{\vAim}{{\cal A}_{i_1\cdots i_{n-1}}}
\newcommand{\vAbi}{\bar{\cal A}^{i_1\cdots i_n}}
\newcommand{\vAbim}{\bar{\cal A}^{i_1\cdots i_{n-1}}}
\newcommand{\htS}{\hat{S}}
\newcommand{\htR}{\hat{R}}
\newcommand{\htI}{\hat{I}}
\newcommand{\htB}{\hat{B}}
\newcommand{\htD}{\hat{D}}
\newcommand{\htV}{\hat{V}}
\newcommand{\cT}{{\cal T}}
\newcommand{\cM}{{\cal M}}
\newcommand{\cMs}{{\cal M}^*}
\newcommand{\vk}{{\bf k}}
\newcommand{\vK}{{\\bf K}}
\newcommand{\vb}{{\textstyle{\bf b}}}
\newcommand{{\vp}}{{\vec p}}
\newcommand{{\vq}}{{\vec q}}
\newcommand{\vQ}{{\vec Q}}
\newcommand{\vx}{{\textstyle{\bf x}}}
\newcommand{\tr}{{{\rm Tr}}}
\newcommand{\beq}{\begin{equation}}
\newcommand{\eeq}[1]{\label{#1} \end{equation}}
\newcommand{\half}{{\textstyle \frac{1}{2} }}
\newcommand{\lton}{\mathrel{\lower.9ex \hbox{$\stackrel{\displaystyle
<}{\sim}$}}}
\newcommand{\gton}{\mathrel{\lower.9ex \hbox{$\stackrel{\displaystyle
>}{\sim}$}}}
\newcommand{\ee}{\end{equation}}
\newcommand{\ben}{\begin{enumerate}}
\newcommand{\een}{\end{enumerate}}
\newcommand{\bit}{\begin{itemize}}
\newcommand{\eit}{\end{itemize}}
\newcommand{\bc}{\begin{center}}
\newcommand{\ec}{\end{center}}
\newcommand{\bea}{\begin{eqnarray}}
\newcommand{\eea}{\end{eqnarray}}
\newcommand{\beqar}{\begin{eqnarray}}
\newcommand{\eeqar}[1]{\label{#1}\end{eqnarray}}
\newcommand{\bra}[1]{\langle {#1}|}
\newcommand{\ket}[1]{|{#1}\rangle}
\newcommand{\norm}[2]{\langle{#1}|{#2}\rangle}
\newcommand{\brac}[3]{\langle{#1}|{#2}|{#3}\rangle}
\newcommand{\hilb}{{\cal H}}
\newcommand{\pleft}{\stackrel{\leftarrow}{\partial}}
\newcommand{\pright}{\stackrel{\rightarrow}{\partial}}

\begin{flushright}
%DRAFT 4/7 10pm
%\vskip .5cm
\end{flushright} \vspace{1cm}

\title{3D Jet Tomography of the Twisted Color Glass Condensate}

\author{A.~Adil}%
\email{azfar@phys.columbia.edu}

\author{M.~Gyulassy}
\email{gyulassy@phys.columbia.edu}

\author{T.~Hirano}
\email{hirano@phys.columbia.edu}

\affiliation{Columbia University, Department of
Physics, 538 West 120-th Street, New York, NY 10027
}%

\date{\today}% It is always \today, today,

\begin{abstract}
  Jet Tomography is proposed as a new test of Color
  Glass Condensate (CGC) initial conditions in non-central $A+A$ collisions.
  The $k_{T}$ factorized CGC formalism is
  used to calculate the rapidity twist in the %$(x_\perp,\eta)$
  reaction plane of  both the bulk low $p_T< 2$ GeV  matter
  as well as the rare 
high $p_T> 6$ GeV partons. Unlike conventional perturbative QCD,
  the initial high $p_{T}$ 
CGC gluons are shown to be twisted even
further away from the beam axis than the 
  the low $p_T$ bulk at high rapidities $|\eta|>2$.  Differential
directed flow $v_{1}(p_{T}>6,|\eta|>2)$ is proposed 
to test this novel high $p_T$ rapidity 
twist predicted by the CGC model.
 \end{abstract}


\maketitle

%%%%%%%%%%%%%%%%%%%%%%%%%%%%%%%%%%%%%%%%%%%%%%%%%%%%%%%%%%%%%%%%%%%%%%%%%%
{\it Introduction:} 
It was pointed out in Ref.\cite{AdilGyulassy:2005BGK} 
that the QCD matter produced in high energy
noncentral $A+A$ nuclear reactions violates ({\em locally})
Bjorken longitudinal
boost invariance in the transverse plane
even if the global rapidity distribution, $dN/d\eta$, is independent 
of $\eta=\sinh^{-1}(p_z/m_\perp)$. The intrinsic
longitudinal
boost non-invariance
occurs even in symmetric $A+A$ reations because
locally in the transverse $\vx_\perp$ plane,
the initial gluon 
density, $\rho_g(\eta,\vx_\perp)=dN_g/d\eta d^2\vx_\perp$, has 
a generic ``trapezoidal'' form in the rapidity variable. 
This peculiar structure arises in noncentral $(b>0)$ collisions
because there is
an asymmetry between
the local  number of interacting projectile and target
nucleons, $\Delta \nu(\vx_\perp; b>0)\sim A^{1/3}$, that can vary
by an order of magnitude in heavy nuclei with the transverse coordinate,
$\vx_\perp$. Since low $p_T$ bulk  matter is known to be produced
proportional to the number of participating nucleons,
the slope of the rapidity trapezoid varies with $\vx_\perp$ as
$d\rho/d\eta\propto \Delta\nu(\vx_\perp; b)/2Y \sim A^{1/3}/\log s$, where
$2Y=\log s$ is the rapidity gap between the projectile and target nuclei.
At infinite energies this slope approaches zero, but at the Relativistic Heavy Ion Collider (RHIC) and the Large Hadron Collider (LHC) it may be large enough to be observable via
3D $(p_\perp,\phi,\eta)$ extensions\cite{AdilGyulassy:2005BGK}
of jet tomography\cite{Gyulassy:2003mc}.

The trapezoidal
rapidity asymmetry has been observed at all energies
in $p+A$ reactions and it was also clearly seen in $D+Au$ reactions at RHIC\cite{Back:2003hx,Arsene:2004cn}.
The data are well reproduced by phenomenological
soft+hard (string+mini-jet) models such 
as \cite{Brodsky:1977de,Andersson:1986gw,Wang:1991ht}.
However, these trapezoidal features are also well reproduced
by gluon saturation
models such as in the Kharzeev-Levin-Nardi (KLN)\cite{Kharzeev:2002ei} implementation of
the
Color Glass Condenstate (CGC) theory\cite{Iancu:2000hn,Blaizot:2004px} 
of low x parton initial conditions. 

However, the phenomenological 
soft+hard  and CGC approaches differ significantly 
in their predictions for moderately 
high $p_T$ partons above $Q_s$. In the former
case, the 
high $p_T$ parton production
is calculated from the collinear factorized on shell $gg\rightarrow
gg$ approximation of pQCD for hard processes. In the KLN/CGC approach
both soft and hard partons
are calculated 
using the $k_T$ factorized off-shell $gg\rightarrow g$ gluon
fusion approximation to QCD parton production\cite{GLR}. 
The advantage of
 collinear factorized approximations is that they employ
experimentally well determined nucleon 
parton structure functions, $xG_N(x,Q^2)$. In addition, corrections
beyond lowest order can be systematically evaluated.
The main disadvantage in  applications to $A+A$
is that possible nonlinear nuclear effects at low x
could significantly modify the assumed
linear relation $G_{A}= AG_{N}$ in different kinematic regimes.
 
In contrast, KLN/CGC predictions are based on the convolution of
unintegrated nuclear 
gluon distributions, $\phi_A(x,k_T^2)$ via the Gribov-Levin-Ryskin (GLR) formula.
This approach  has the advantage
of including a nonlinear gluon evolution as $x\rightarrow 0$
via a single gluon saturation scale $Q_s(x,A)$ that can be determined in the weak coupling but strong field approximation.
The main disadvantage of this approach
is the uncertainty related to how to
take the 
$A\rightarrow 1$ nucleon limit,
$\phi_N(x,k_T^2)$. This limit is important phenomenologically
because nuclear modification is measured relative
to the $N+N$ baseline. In addition, near the nuclear surface
the participant nucleon density decreases rapidly and $Q_s$ is driven
below 1 GeV where the weak coupling strong field 
approximations become suspect.
Since the surface regions in finite A+A contribute significantly
($\sim 10-20$ \%) to the global $dN/dy$, 
the specific 
implementation of the low $A$ limit is important to devise 
experimental tests sensitive to this limit of CGC.
Unlike the integrated parton distributions, 
there is no general consensus yet
on the form of $\phi_N$ \cite{Andersson:2002cf}. 
Additional theoretical uncertainty is associated with the unknown
applicability range
 of the first order GLR $k_T$ factorization
formulation\cite{GLR} as compared to 
proposed higher order nonlinear generalizations\cite{Nikolaev:2004cu}.

At RHIC the strongest support for the KLN/CGC approach
is its remarkable 
ability to reproduce the extensive systematics of
the energy and
nuclear size dependence of the global $p_T$ integrated
$dN_{ch}/dy$. This results from a specific 
dependence of the saturation scale, $Q_s$, on $\sqrt{s}$ and $A$.
In contrast, phenomenological soft-hard models such as HIJING
fail to reproduce the systematics
because  the separation scale, $p_0\sim 2 $ GeV, 
between soft and hard parton production
was assumed to be independent of those variables.
It is an important open question of how high in $p_T$
can the $k_T$-factorized KLN/CGC approximation 
be pushed in specifying the $A+A$ initial conditions
versus how low in $p_T$ can the conventional collinear factorized
approximation be pushed. 
Both experimental and theoretical control over the initial conditions
in $A+A$ at RHIC are essential to strengthen the current case
for the discovery of new forms of matter, the strongly coupled Quark Gluon Plasma (sQGP) and CGC,  at RHIC\cite{Gyulassy:2004zy,
Adcox:2004mh,HiranoGyulassy:2005}. 

\begin{figure}[t!]
\begin{center}
%\hspace*{-.6in}
\epsfig{file=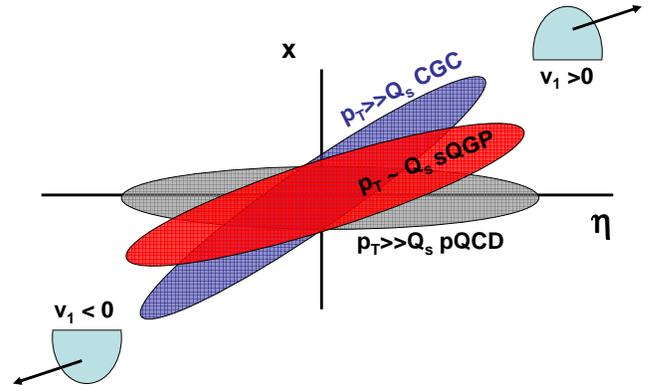, width=2in,angle=270}
\caption{(Color Online) Illustration of the initially twisted 
sQGP gluon density\cite{AdilGyulassy:2005BGK} relative to the beam axis
in the $(x,\eta)$ reaction plane. CGC\cite{Kharzeev:2002ei},
BGK\cite{Brodsky:1977de}, and HIJING\cite{Wang:1991ht}
predict this generic  low $p_T \lton Q_s$ 
locally boost non-invariant structure. Also shown are 
the relative rotations of the high $p_{T}\gg Q_s$ 
jet partons in the $k_T$ factorized CGC model as well as
conventional collinear factored
pQCD. Jet quenching through the sQGP leads to opposite 
sign first azimuthal harmonic moment,
$v_1(p_T\gg Q_s,\eta)$ in the two approaches. The projectile and spectator
nuclei are indicated by half circles together with the sign convention
of low $p_T$ directed flow $v_1$.
}
\label{CGCBulkFig}
\end{center}
\end{figure}

In this letter a  new jet tomographic approach is  proposed
based on extending the discussion
in Ref.\cite{AdilGyulassy:2005BGK} to KLN/CGC initial conditions
of the sQGP bulk as well its extrapolation to high $p_T\gg Q_s$ jet partons.
The idea is to exploit the difference between
the geometric distributions of jets 
relative to that of the bulk matter as illustrated in Fig.\ref{CGCBulkFig}.
We focus on the  predicted azimuthal dependence
of the jet quenching pattern,  $R_{AA}(\eta,\phi,p_\perp; b>0)$,
and long range rapidity correlations induced
by the generic intrinsic rapidity
twist of the bulk matter.

At midrapidity, $\eta=0$, the elliptic asymmetry of the 
reaction geometry in noncentral $(b>0)$ in $A+A$ 
reactions leads to a well known elliptic asymmetry in jet quenching.
However, the rapidity twist has no observable effect
at mid-rapidity. At positive rapdities
$\eta>2$ , on the other hand the rapidity twist of the bulk shifts the center
of mass away from $x=0$, while at negative rapidities that
shift has opposite sign. In conventional pQCD, collinear
factorized $gg\rightarrow gg$ predicts 
a jet distribution that is proportional to the
local {\em binary} collision density, $\sigma_{NN}T_{B}(r_+)T_{A}(r_-)$, where
$r_\pm=\sqrt{(x\pm\frac{b}{2})^{2}+y^{2}}$, and
$T_A(\vx)$ is the Glauber nuclear profile 
function\cite{AdilGyulassy:2005BGK}.  The jets are therefore produced
symmetrically about 
$\vx_\perp=0$ at all $\eta$. The collinear factorized jet density 
therefore has no rapidity twist as illustrated by the grey ellipse in
Fig.\ref{CGCBulkFig}

The bulk matter density is effectively rotated away from
the beam axis because the bulk density 
varies approximately as
$\{(Y-\eta)T_A(r_+)+ (Y+\eta)T_A(r_-)\}/2Y$, which is not reflection symmetric
about $\vx_\perp=0$ away from midrapidity.
At $p_T < Q_s$ the CGC model produces approximately the same rapidity twist
as wounded nucleon string models since this is a direct 
consequence of local
participant versus binary collision scaling of the bulk.
However, due to the nonlinear equation determining the local
saturation scale $Q_s(\vx_\perp,x)$ the bulk density surface region
can sharpen significantly over conventional
Wood-Saxon geometry included in $T_A(\vx_\perp)$. This surprising
change of the bulk surface geometry in the KLN implementation
of CGC 
was first pointed out by Hirano and Nara\cite{Hirano:2004rs}.

However, as we show below, at higher $p_T$ the CGC predicts even
greater rapdity twist away from the beam axis than the bulk as shown
below and illustrated in Fig.\ref{CGCBulkFig}.  In this paper, we
explore some tomographic consequences of using the $k_{T}$ factorized
formalism and CGC type unintegrated gluon distributions to produce
both the bulk and jet matter in the reaction.  
%We predict an increased
%intrinsic rotation of the initially produced high $p_{T}$ matter
%relative to the low $p_{T}$ bulk.  
This anomalous rapidity twist
effect is opposite to that discussed in \cite{AdilGyulassy:2005BGK}.  This
effect occurs because the different nuclei are probed at 
asymmetric Bjorken momentum fractions while producing high $p_{T}$
matter.

{\it The Local Gluon Distribution:} The local generalization of
$k_{T}$-factorization
GLR formula\cite{GLR}
 used by KLN\cite{Kharzeev:2002ei} and Hirano and Nara
\cite{Hirano:2004rs} is given by
\begin{eqnarray}
  \frac{dN_{g}}{dp_{T}d^{2}x_{T}d\eta}=\frac{4\pi}{C_{F}}\frac{\alpha_{s}(p_T^{2})}{p_{T}}
\int^{p_T} d^{2}k_{T} \times \qquad \qquad \quad \nonumber \\
 \phi_{A}(x_{1},(\frac{\vec{k}_{T}+\vec{p}_{T}}{2})^{2};\vec{x}_{T}) \phi_{B}(x_{2},(\frac{\vec{k}_{T}-\vec{p}_{T}}{2})^{2};\vec{x}_{T}).
 \label{eqn:ktfac}
\end{eqnarray}
$C_{F}=\frac{N_{C}^{2}-1}{2N_{C}}$ %is the SU($\mathrm{N_{C}}$) Casimir 
and the collinear momentum fractions are given by kinematics,
$x_{1,2}=p_{T}\exp(\pm \eta)/\sqrt{s}$.  The QCD coupling,
$\alpha_{s}$, is evaluated at $p_{T}^{2}$ and regulated at low $p_T$
by imposing a maximum value $\alpha_{\textrm{max}}=0.5$. 
 Note that we use
$\eta$ to denote the rapidity rather than the pseudo rapidity as in
\cite{AdilGyulassy:2005BGK}.

$\phi_{A,B}$ are the unintegrated gluon distributions which, in
principle, possess a Bjorken $x$ dependence determined by nonlinear
evolution equations of the CGC theory\cite{Iancu:2000hn,Blaizot:2004px}
 and their $k_{T}$ dependence is fixed by 
a characteristic saturation momentum, $Q_{s}(x)$.
In the McLerran-Venugopalan approach
\cite{McLerrVenu:1994} the gluon distribution is suppressed below the
saturation scale $\phi_A \sim\log({Q_{S}^{2}}/{k_{T}^{2}})$ compared to
the perturbative form $\phi_A\sim {k_{T}^{-2}}$.  We use
a  parameterization
similar to the KLN model approach as used  in \cite{Hirano:2004rs}.
However, we use (for numerical convenience) the following Lorentzian form of
$\phi_{A,B}$ for all values of $k_{T}$.
\begin{equation}
\phi_{A}(x,\vec{k}_{T};\vec{x}_{T})=\frac{\kappa }{\alpha_{s}(Q_{s,A}^{2})}\frac{Q_{s,A}^{2}}{k_{T}^{2}+Q_{s,A}^{2}+\Lambda^{2}}.
\label{eqn:phiab}
\end{equation}
The momentum scale $\Lambda=0.2$ GeV is a regulator for the high
rapidity $y>4.5$ region as in \cite{Hirano:2004rs}.  The constant
$\kappa\sim 0.5$ is a parameter set to reproduce $dN_{g}/d\eta \sim1000$
at midrapidity central collisions.  The transverse coordinate
dependence is implicit in the saturation momentum determined
numerically for each nucleus.
\begin{equation}
Q_{s,A}^{2}(x,\vec{x}_{T})=\frac{2\pi^{2}}{C_{F}}\alpha_{s}(Q_{s,A}^{2})xG_{\textrm{nuc}}(x,Q_{s,A}^{2})T_{A}(\vec{x}_{T}),
\label{eqn:qsat}
\end{equation}
where $T_{A}$ is the Glauber profile of nucleus $A$.  We use standard
diffuse
Woods-Saxon profiles.  The projectile and target nucleii are set up
such that the spectator $v_{1}$ is
positive at forward rapidity.

The  KLN parametrization  is used for the nucleonic gluon distribution.
\begin{equation}
xG_{\textrm{nuc}}(x,Q^{2})=K\log(\frac{Q^{2}+\Lambda^{2}}{\Lambda_{QCD}^{2}})x^{-\lambda}(1-x)^{n}
\label{eqn:nucglu}
\end{equation}
The momentum scales $\Lambda$ and $\Lambda_{\textrm{QCD}}$ are set to
0.2 GeV.  The $x^{-\lambda}$ term accounts for the rapid growth
of small x gluons while the factor of
$(1-x)^{n}$ was introduced in KLN to account qualitatively
for the rapid depletion of gluons as
$x\rightarrow 1$ outside the small x framework
of the CGC model.  As in the KLN approach, we set $\lambda = 0.2$ and
$n =4$.  $K\sim1.35$ is used to set $\langle Q_{s}^{2}(x=0.01)\rangle
\sim 2$ GeV$^{2}$ for central collisions at midrapidity.

\begin{figure}
\centering
 \epsfig{file=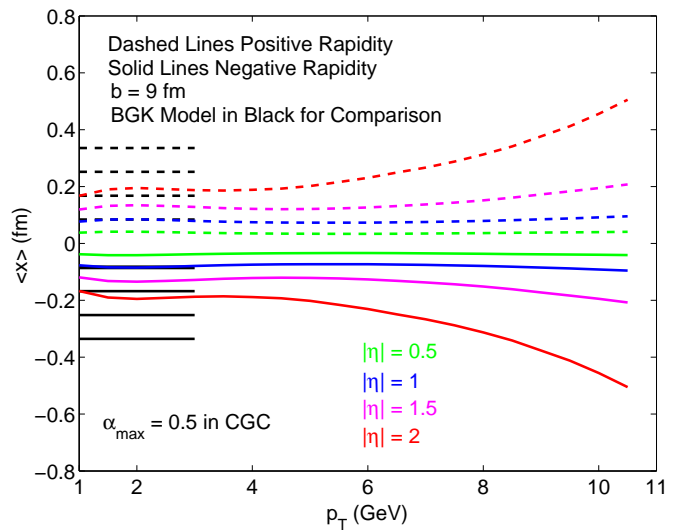,width=3.5in,angle=0}
  \caption{(Color Online) The figure shows the average transverse spatial coordinate $\langle x \rangle$ for produced gluonic matter in the Brodksy-Gunion-Kuhn \cite{Brodsky:1977de} type twist model as in \cite{AdilGyulassy:2005BGK}, as well as in the current CGC model.  Notice that, for BGK, there is no dependence on $p_{T}$ other than the assumed lack of twist for matter with $p_{T} > 3$ GeV.  The CGC model, however, has a pronounced greater twist for higher $p_{T}>6$ GeV.
}
  \label{BGKandCGCAvx}
\end{figure}

Once the distribution shown in Eq. \ref{eqn:ktfac} is evaluated, we are ready to investigate the transverse plane dependence (and hence the intrinsic spatial twist) of the produced gluons as a function of $p_{T}$ and $\eta$.  We measure this shift of material away from the centre of the transverse reaction plane by calculating the average horizontal transverse coordinate.
\begin{equation}
\langle x \rangle(p_{T},\eta)=\frac{\int d^{2}x_{T} x dN_{g}/dp_{T}d^{2}x_{T}d\eta}{\int d^{2}x_{T} dN_{g}/dp_{T}d^{2}x_{T}d\eta}
\label{eqn:avx}
\end{equation}
Hereafter, $x$ is the transverse coordinate in the direction of the
reaction plane, $\vec{b}$. Positive $x$ points toward the projectile
$\eta=Y$ spectator displacement .  Fig. \ref{BGKandCGCAvx} shows
$\langle x \rangle$ as a function of $p_{T}$ and $\eta$ for the hybrid
Brodsky-Gunion-Kuhn (BGK) \cite{Brodsky:1977de} participant and binary
jet production model (as used in \cite{AdilGyulassy:2005BGK}) as well
as for the current CGC model (for $b=9$ fm).  The rapidity twist of the
BGK model is seen by the increasing $\langle x \rangle$ as $\eta $
increases for the bulk $p_{T}\leq3$ GeV matter.
%midrapidity while the jet matter is wholly undisturbed and remains
%untwisted at all rapidites.

%The situation is changed in the CGC model.  
While the bulk $\langle x\rangle$ of the CGC model is similar to the bulk BGK
shifts, the high $p_T$ shifts behave oppositely. In the two component BGK
approach $\langle x\rangle = 0$ for high $p_T$.
%source then there is no need for a hybrid model that patches together
%different soft particle production and jet production processes.  One
%just simply achieves an expression as in Eq. \ref{eqn:ktfac} for the
%differential distribution of gluons produced.  
We show in Fig.
\ref{BGKandCGCAvx} that the rapidity twist ($d\langle x\rangle/d\eta$)
increases at high $p_{T}>6$  GeV.
% asymmetry is very similar to
%the result achieved in the BGK model.  
%The interesting deviation from
%the previous model is that the high $p_{T}$ matter, far from being
%untwisted, is actually twisted to a greater extent than the soft
%produced matter.

{\it Tomography and the Inverse Twist:} 
Jet tomographic analysis uses the attenuation of jet matter while passing through the bulk in order to gain information about the density profile of the bulk \cite{Gyulassy:2003mc}.  The observable used most commonly in tomographic analysis is the nuclear modification factor, $R_{AA}$,
which measures the deviation of the produced nucleus-nucleus spectrum, if any, from a simple binary scaled p-p spectrum.  The twist effect investigated in the previous section can be observed by looking at the $R_{AA}(p_{T},\eta,\phi)$ of jets in the transverse plane.

The azimuthal dependence of $R_{AA}$ will change as a function of $\eta$ for a given $p_{T}$ jet due to the differing twist of the jet distribution over $\eta$. 
\begin{figure}
\centering
 \epsfig{file=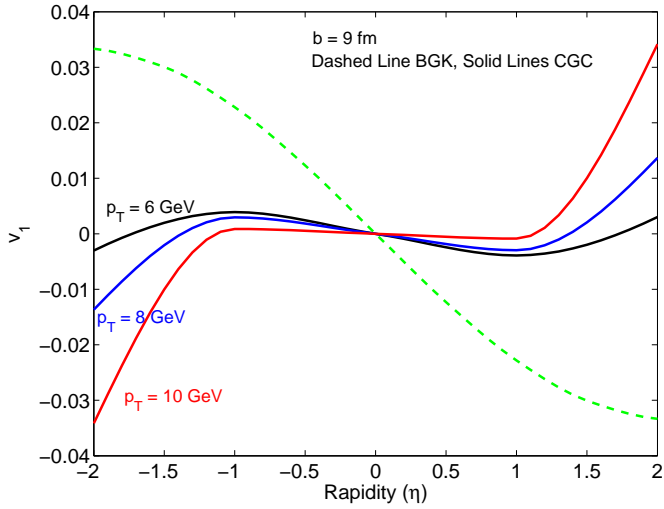,width=3.5in,angle=0}
  \caption{(Color Online) The directed flow $v_{1}$ as a function of $\eta$ for different $p_{T}$ at $b=9$ fm. Both the CGC model and BGK model are given for comparison.}
    \label{v1}
\end{figure}
In order to calculate the $R_{AA}(p_{T},\eta,\phi)$ we use the geometric model of Drees, Feng and Jia \cite{Drees:2003zh}.  The nuclear modification factor is obtained by,
\begin{equation}
R_{AA}(p_{T},\eta,\phi)=\frac{\int d^{2}x_{T}e^{-\mu \chi(\vec{x}_{T},\phi,\eta)}\frac{dN_{g}}{dp_{T}d^{2}x_{T}d\eta}(p_{T},\eta)}{\int d^{2}x_{T}\frac{dN_{g}}{dp_{T}d^{2}x_{T}d\eta}(p_{T},\eta)}.
\label{eqn:gRAA}
\end{equation}
$\mu = 0.04$  is the parameter used to set $R_{AA}(\eta=0,b=0)\sim0.25$.  Opacity, $\chi$, is the line integral over the bulk distribution that is the cause of the attenuation experienced by the jet, calculated as in \cite{Drees:2003zh}.
The length dependence of opacity is characteristic of 
radiative parton energy loss in Bjorken expanding matter.

We can use Eq. \ref{eqn:gRAA} to probe the twist.  The anti-twist
effect can most easily observed via the first azimuthal fourier moment
of $R_{AA}(p_{T},\eta,\phi)$, the directed flow $v_{1}$.  Fig.
\ref{v1} shows $v_{1}$ as a function of $\eta$ for different values of
the $p_{T}$.  Note that for all $p_{T}$ values there exists a 
rapidity at which the directed flow flips sign.
This flip occurs at lower values of the rapidity for higher values of
the $p_{T}$.  The sign of $v_1$ is positive in CGC at high rapidity
in the same direction as the projective spectator. The change in sign is
a novel prediction using the KLN/CGC model.
In conventional factorized QCD jet production, the high $p_T$
$v_1$ is negative and in the same direction as the low $p_T$ 
bulk directed flow but  increasing with $p_T$ as in hydrodynamics.
Both models exhibit long range rapidity
anticorrelations of the $v_1$. This anticorrelation
may make it easier to measure the sign of $v_1$ and test up to how
high $p_T$ can the KLN/CGC model describe  
initial conditions at RHIC.

\begin{acknowledgments}
Discussions with J. Harris, W. Horowitz, D. Kharzeev, L. McLerran, 
I. Vitev, X.N. Wang and Nu Xu
are gratefully acknowledged.
  This work is supported in part by the United States
Department of Energy
under Grants   No. DE-FG02-93ER40764.
\end{acknowledgments}

%\begin{thebibliography}{80}

%\end{thebibliography}

\end{document}